\documentclass[useAMS]{mn2e}
\usepackage{graphicx}

\def\arcmin{\tt '}

\def\simgt{\ {\raise-.5ex\hbox{$\buildrel>\over\sim$}}\ }
\def\I{\'\i}
\def\cd{cd$^{-1}$\,}

\begin{document}

\title[Asteroseismology of $\nu$~Eridani: photometry]
{Asteroseismology of the $\beta$ Cephei star $\nu$~Eridani:
photometric observations and pulsational frequency analysis}
\author[G. Handler et al.]
  {G. Handler$^{1}$, R. R. Shobbrook$^{2}$, M. Jerzykiewicz$^{3}$, K.
Krisciunas$^{4,5}$, \and T. Tshenye$^{6}$, E. Rodr\I guez$^{7}$, V.  
Costa$^{7}$, A.-Y. Zhou$^{8}$, R. Medupe$^{6, 9}$, \and W. M.
Phorah$^{6}$, R. Garrido$^{7}$, P. J. Amado$^{7}$, M. Papar\'o$^{10}$, D. 
Zsuffa$^{10}$, \and L. Ramokgali$^{6}$, R. Crowe$^{11}$, N. Purves$^{11}$, 
R. Avila$^{11}$, R. Knight$^{11}$,\and E. Brassfield$^{11}$, P. M. 
Kilmartin$^{12}$, P. L. Cottrell$^{12}$
\and \\
$^1$ Institut f\"ur Astronomie, Universit\"at Wien, T\"urkenschanzstrasse
17, A-1180 Wien, Austria\\
$^{2}$ Visiting Fellow, Australian National University, Canberra, ACT,
Australia\\
$^{3}$ Wroclaw University Observatory, ul. Kopernika 11, 51-622 Wroclaw, 
Poland\\
$^{4}$ Cerro Tololo Inter-American Observatory, National Optical Astronomy 
Observatory, Casilla 603, La Serena, Chile\\
$^{5}$ Las Campanas Observatory, Casilla 601, La Serena, Chile\\
$^{6}$ Theoretical Astrophysics Programme, University of the North-West,
Private Bag X2046, Mmabatho 2735, South Africa\\
$^{7}$ Instituto de Astrofisica de Andalucia, C.S.I.C., Apdo. 3004, 18080 
Granada, Spain\\
$^{8}$ National Astronomical Observatories, Chinese Academy of Sciences, 
Beijing 100012, China\\
$^9$ South African Astronomical Observatory, P.O. Box 9, Observatory 7935,
South Africa\\
$^{10}$ Konkoly Observatory, Box 67, H-1525 Budapest XII, Hungary\\
$^{11}$ Department of Physics and Astronomy, University of Hawaii - Hilo, 
200 West Kawili Street, Hilo, Hawaii, 96720-4091, USA\\
$^{12}$ Department of Physics and Astronomy, University of Canterbury,
Christchurch, New Zealand}

\date{Accepted 2003 July 17.
  Received 2003 August 13;
  in original form 2003 September 10}
\maketitle
\begin{abstract}

We undertook a multisite photometric campaign for the $\beta$ Cephei star
$\nu$~Eridani. More than 600 hours of differential photoelectric $uvyV$
photometry were obtained with 11 telescopes during 148 clear nights.

The frequency analysis of our measurements shows that the variability of
$\nu$~Eri can be decomposed into 23 sinusoidal components, eight of which
correspond to independent pulsation frequencies between 5--8 \cd. Some of
these are arranged in multiplets, which suggests rotational $m$-mode
splitting of nonradial pulsation modes as the cause. If so, the rotation
period of the star must be between 30 -- 60 d.

One of the signals in the light curves of $\nu$~Eri has a very low
frequency of 0.432 \cd. It can be a high-order combination frequency or,
more likely, an independent pulsation mode. In the latter case $\nu$~Eri
would be both a $\beta$ Cephei star and a slowly pulsating B (SPB) star.

The photometric amplitudes of the individual pulsation modes of $\nu$~Eri
appear to have increased by about 20 per cent over the last 40 years. So
do the amplitudes of the dominant combination frequencies of the star.
Among the latter, we only could identify sum frequencies with certainty,
not difference frequencies, which suggests that neither light-curve
distortion in its simplest form nor resonant mode coupling are their
single cause.

One of our comparison stars, $\mu$~Eridani, turned out to be variable with
a dominant time scale of 1.62 d. We believe that it is either an SPB star
just leaving its instability strip or that its variations are of
rotational origin.

\end{abstract}


{\section{Introduction}}

Lengthy multisite observations of multiperiodically pulsating stars have
become a standard tool in variable star research. The benefit of such
efforts are long, uninterrupted time series of the variations of the
target stars, which are necessary to resolve complicated pulsational
spectra. The more individual variations are present in a given pulsator,
the more one can learn about its interior by modelling the observed mode
spectra. This technique is called asteroseismology.

The most extensive observational efforts for asteroseismology have been
performed with dedicated telescope networks. For instance, the Whole Earth
Telescope (Nather et al. 1990) has already observed 40 individual targets,
more than half of which are pulsating white dwarf stars, the others being
pulsating sdB stars, rapidly oscillating Ap stars, cataclysmic variables
etc. The Delta Scuti Network (e.g.\ Zima et al.\ 2002) has studied 10
different objects during 23 campaigns and acquired a total of more than
1000 hours of measurement for some $\delta$ Scuti pulsators.

The very first coordinated multisite observations were obtained as early
as 1956, on the $\beta$ Cephei star 12 (DD) Lacertae (de Jager 1963). This
effort even included both spectroscopic and multicolour photometric
measurements. In more recent times, however, $\beta$ Cephei stars were
(aside from a large campaign for BW Vul, Sterken et al. 1986) rarely the
targets of extended observing campaigns. The reason may be the sparse
frequency spectra of $\beta$ Cephei stars compared to pulsating white
dwarfs or $\delta$ Scuti stars. For many years, the record holder was 12
(DD) Lac with five known independent modes of pulsation (Jerzykiewicz
1978), which was recently superseded by the six modes of V836 Cen (Aerts
et al. 2003).

However, the apparent paucity of frequencies in the mode spectra of
$\beta$ Cephei stars may be questioned. Experience with $\delta$ Scuti
stars has shown that the more the detection level for periodic light
variations is pushed down, the more pulsation modes are detected (e.g.\
see the sequence of papers by Handler et al.\ 1996, 1997 and 2000). In
fact, most of the pulsation modes of these stars have light amplitudes
around or below 1 mmag, an amplitude quite easily detected with the large
data sets of 2 to 3 mmag precision differential photometry obtained during
these campaigns. As the pulsational driving of the $\beta$ Cephei stars
(Moskalik \& Dziembowski 1992) is based on essentially the same mechanism
(the $\kappa$ mechanism) as that of the $\delta$ Scuti stars, just
operating on heavier chemical elements, it can be suspected that many
low-amplitude modes are also excited in $\beta$ Cephei stars but have not
yet been detected simply because of a lack of suitable data. Indeed, this
idea is supported by recent high-quality observations (Stankov et al.
2002, Cuypers et al. 2002, Handler et al. 2003).

Besides the detection of many pulsation modes, another necessary
ingredient for asteroseismology is the correct identification of these
modes with their pulsational quantum numbers, $k$, the radial overtone of
the mode, the spherical degree $\ell$ and the azimuthal order $m$. For
pulsating stars whose frequency spectra do not show any obvious
regularities caused by rotationally split modes or consecutive radial
overtones the use of mode identification methods is required. This may for
instance be spectroscopic diagnostics from line profile variations or
photometric colour amplitude ratios and phase shifts. Unfortunately, such
methods may not always yield unambiguous results (e.g.\ see Balona 2000).
However, Handler et al.\ (2003) recently showed that mode identification
from photometric colour amplitudes works well for slowly rotating $\beta$
Cephei stars and they estimated that a relative accuracy of 3 per cent in 
the amplitude determinations are sufficient to achieve an unambiguous
determination of $\ell$.

Consequently, the $\beta$ Cephei stars are indeed suitable for
asteroseismic studies. If successful, many interesting astrophysical
results can be expected. For instance, angular momentum transport in these
stars can be studied. The frequencies of some pulsation modes of $\beta$
Cephei stars are sensitive to the amount of convective core overshooting
(Dziembowski \& Pamyatnykh 1991). Deviations from the rotational frequency
splitting of nonradial mode multiplets can be due to interior magnetic
field structure of those stars (Dziembowski \& Jerzykiewicz 2003). Once
the interior structures of several $\beta$ Cephei stars in various phases
of their evolution are determined, main-sequence stellar evolution
calculations can be calibrated and more accurately extrapolated to the
supernova stage, which can in turn constrain spectral and chemical
evolution theories of galaxies.

Hence it is justified to devote large observational efforts to $\beta$
Cephei stars that seem suitable for asteroseismology. The selection of a
good candidate is one of the most important prerequisites for such a
study. For the present work, our choice was $\nu$~Eri (HD 29248,
$V=3.92$). Its mode spectrum reveals high asteroseismic interest: four
pulsation frequencies were known, a singlet and an equally spaced triplet
(Kubiak 1980, Cuypers \& Goossens 1981). The singlet has been suggested to
be a radial mode, and the triplet is consistent with a dipole (Aerts,
Waelkens \& de Pauw 1994, Heynderickx, Waelkens \& Smeyers 1994).

If this triplet contained at least two rotationally split $m$-components
of a mode, $\nu$~Eri would also be a slow rotator, a hypothesis supported by
its measured $v \sin i$ (the most recent determination being 20 km/s, Abt,
Levato \& Grosso 2002). This is important because the adverse effects of
rotational mode coupling (see Pamyatnykh 2003 or Daszy{\'n}ska-Daszkiewicz
et al.\ 2002) in a subsequent theoretical analysis would be diminished.
Finally, $\nu$~Eri is a bright equatorial star, so it can be observed from
both hemispheres with photometric and high-resolution spectroscopic
instruments.

We therefore organised a multisite campaign for $\nu$~Eri, applying both
observing methods mentioned above (Handler \& Aerts 2002). In the
following, we will report the results from the photometric measurements.  
The analysis of the spectroscopy, pulsational mode identification and
seismic modelling of the identified oscillations will be the subject of
future papers.

\vspace{4mm}

\section{Observations and reductions}

Our photometric observations were carried out with 11 different telescopes
and photometers at 10 observatories on 5 different continents; they are
summarised in Table 1. In most cases, single-channel differential
photometry was acquired through the Str\"omgren $uvy$ filters. However, at
Sierra Nevada Observatory (OSN) a simultaneous $uvby$ photometer was used,
so we included the $b$ filter as well, and at the four observatories where
no Str\"omgren filters were available we used Johnson $V$. Some
measurements through the H$_{\beta}$ filters were also obtained at OSN.  
The total time base line spanned by our measurements is 157.9 d.

\begin{table*}
\caption[]{Log of the photometric measurements of $\nu$~Eri. Observatories 
are ordered according to geographical longitude. Sites that acquired $V$ 
measurements only are marked with asterisks.}
\begin{center}
\begin{tabular}{lcccccl}
\hline
% add long title, longitude, latitude
Observatory & Longitude & Latitude & Telescope & \multicolumn{2}{c}{Amount 
of data} & Observer(s)\\
& & & & Nights & h & \\
\hline
Sierra Nevada Observatory & $-$3\degr 23\arcmin & +37\degr 04\arcmin & 0.9m 
& 18 & 53.59 & ER,\,VC,\,RG,\,PJA\\
Cerro Tololo Interamerican Observatory & $-$70\degr 49\arcmin & $-$30\degr 
09\arcmin & 0.6m & 8 & 43.19 & KK\\
Fairborn Observatory & $-$110\degr 42\arcmin & +31\degr 23\arcmin & 0.75m 
APT & 24 & 114.54 & $--$\\
Lowell Observatory & $-$111\degr 40\arcmin & +35\degr 12\arcmin & 0.5m & 10 
& 46.01 & MJ\\
Mauna Kea Observatory$^*$ & $-$155\degr 28\arcmin & +19\degr 50\arcmin & 
0.6m & 4 & 7.78 & RC,\,NP,\,RA,\,RK,\,EB\\
Mount John University Observatory & +170\degr 28\arcmin & $-$43\degr 
59\arcmin & 0.6m & 1 & 3.83 & PMK\\
Siding Spring Observatory & +149\degr 04\arcmin & $-$31\degr 16\arcmin & 
0.6m & 31 & 117.70 & RRS\\
Xing-Long Observatory$^*$ & +117\degr 35\arcmin & +40\degr 24\arcmin & 
0.85m & 3 & 15.72 & AYZ\\
South African Astronomical Observatory &+20\degr 49\arcmin & 
$-$32\degr 22\arcmin & 0.5m & 37 & 151.82 & GH,\,TT,\,RM,\,WP,\,LR\\
South African Astronomical Observatory & +20\degr 49\arcmin &
$-$32\degr 22\arcmin & 0.75m & 7 & 39.31 & GH\\
Piszkestet\"o Observatory$^*$ & +19\degr 54\arcmin & +47\degr 55\arcmin & 
0.5m & 5 & 11.86 & MP,\,DZ\\
\hline
Total & & & & 148 & 605.35 \\
\hline
\end{tabular}
\end{center}
\end{table*}

We chose two comparison stars for $\nu$~Eri: $\mu$~Eri (HD 30211, B5IV,
$V=4.00$) and $\xi$~Eri (HD 27861, A2V, $V=5.17$). Another check star, HD
29227 (B7 III, $V=6.34$) was also monitored at OSN. We note that $\mu$~Eri
was the single comparison star in all published extensive photometric
studies of $\nu$~Eri (van Hoof 1961, Kubiak \& Seggewiss 1991) and that
its HIPPARCOS photometric data (ESA 1997) imply some slow variability
(Koen \& Eyer 2002). The star is also a spectroscopic binary ($P_{\rm
orb}=7.35890$\,d, $e=0.26$, Hill 1969). In the hope that we could also
understand the variability of $\mu$~Eri with our multisite observations,
and hoping to use that knowledge in re-analyses of the published data of
$\nu$~Eri, we retained $\mu$~Eri as a comparison star. During data
reduction, we took care that the variations of $\mu$~Eri would not
influence the results on our primary target.

Data reduction was therefore started by correcting for coincidence losses,
sky background and extinction. Nightly extinction coefficients were
determined with the Bouguer method from the $\xi$~Eri measurements only;
second-order colour extinction coefficients were also determined. We
then calculated differential magnitudes between the comparison stars (in
the sense $\mu$~Eri - $\xi$~Eri). Heliocentrically corrected versions of
these time series were set aside for later analysis of the variability of
$\mu$~Eri, to which we will return later.

The nightly ($\mu$~Eri - $\xi$~Eri) differential magnitudes were fitted
with low-order polynomials ($n<4$). The residuals of the nondifferential
$\mu$~Eri magnitudes with respect to that fit were combined with the
$\xi$~Eri data and were assumed to reflect the effects of transparency and
detector sensitivity changes only. Consequently, these combined time
series were binned into intervals that would allow good compensation for
the above mentioned nonintrinsic variations in the target star time series
and were subtracted from the measurements of $\nu$~Eri. The binning
minimises the introduction of noise in the differential light curve of the
target.

The timings for this differential light curve were heliocentrically
corrected as the next step. Finally, the photometric zeropoints of the
different instruments, that may be different because of the different
colours of $\nu$~Eri and $\xi$~Eri, were examined at times of overlap with
a different site and adjusted if necessary.  Measurements in the
Str\"omgren $y$ and Johnson $V$ filters were treated as equivalent and
analysed together due to the same effective wavelength of these filters.
The resulting final combined time series was subjected to frequency
analysis;  we show some of our light curves of $\nu$~Eri in Fig.\ 1. In
the end, we had more than 3000 measurements in each filter with accuracies
of 3.7 ($u$ filter), 3.0 ($v$ filter) and 3.0 mmag ($y/V$ filters) per
data point available.

\begin{figure*}
\includegraphics[width=184mm,viewport=-45 00 560 655]{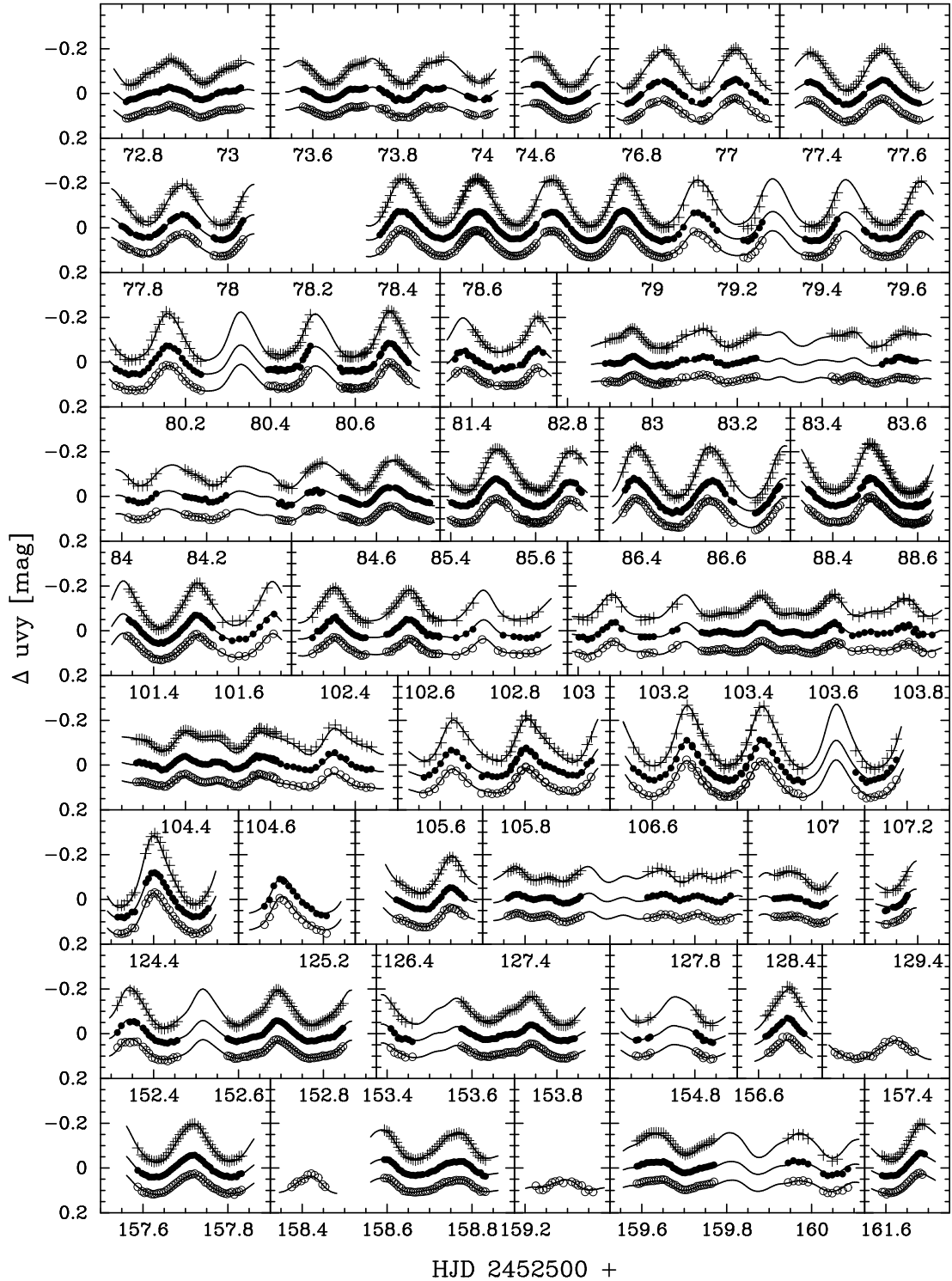}
\caption[]{Some light curves of $\nu$~Eri. Plus signs are data in the 
Str\"omgren $u$ filter, filled circles are our $v$ measurements and open 
circles represent Str\"omgren $y$ and Johnson $V$ data. The full line is a
fit composed of all the periodicities detected in the light curves (Table 2). 
The amount of data shown here is about half the total.} 
\end{figure*}

\section{Frequency analysis}

\subsection{The program star}

Our frequency analyses were mainly performed with the program {\tt PERIOD
98} (Sperl 1998). This package applies single-frequency power spectrum
analysis and simultaneous multi-frequency sine-wave fitting. It also
includes advanced options, such as the calculation of optimal light-curve
fits for multiperiodic signals including harmonic, combination, and
equally spaced frequencies. As will be demonstrated later, our analysis
requires some of these features.

We started by computing the Fourier spectral window of the final light
curves in each of the filters. It was calculated as the Fourier transform
of a single noise-free sinusoid with a frequency of 5.7633 \cd (the
strongest pulsational signal of $\nu$~Eri) and an amplitude of 36 mmag
sampled in the same way as were our measurements. The upper panel of Fig.\
2 contains the result for the combined $y$ and $V$ data. Any alias
structures that would potentially mislead us into incorrect frequency
determinations are quite low in amplitude due to our multisite coverage.

\begin{figure*}
\includegraphics[width=184mm,viewport=-65 00 525 615]{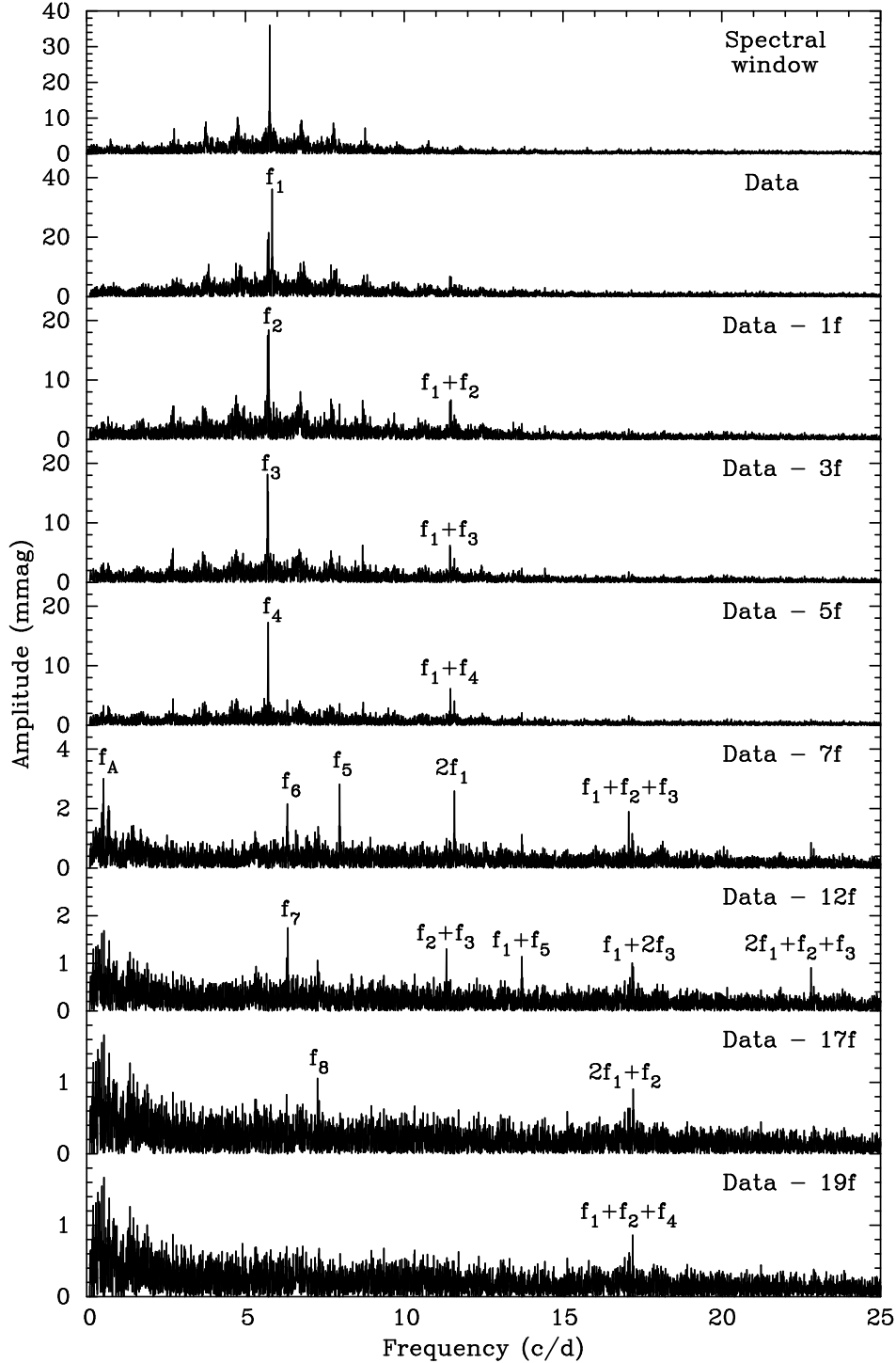}
\caption[]{Amplitude spectra of $\nu$~Eri. The uppermost panel shows 
the spectral window of the data, followed by the periodogram of the data. 
Successive prewhitening steps are shown in the following panels; note 
their different ordinate scales. See text for details.} 
\end{figure*}

We proceeded by computing the amplitude spectra of the data itself (second
panel of Fig.\ 2). The signal designated $f_1$ dominates, but some
additional structures not compatible with the spectral window sidelobes
are also present. Consequently, we prewhitened this signal by subtracting
a synthetic sinusoidal light curve with a frequency, amplitude and phase
that yielded the smallest possible residual variance, and computed the
amplitude spectrum of the residual light curve (third panel of Fig.\ 2).

This resulted in the detection of a second signal ($f_2$) and of another
variation at the sum frequency of the two previously detected. We then
prewhitened a three-frequency fit from the data using the same
optimisation method as before and fixed the combination term to the exact
sum of the two independent frequencies. We continued this procedure
(further panels of Fig.\ 2) until no significant peaks were left in the
residual amplitude spectrum.

We consider an independent peak statistically significant if it exceeds an
amplitude signal-to-noise ratio of 4 in the periodogram; combination
signals must satisfy $S/N>3.5$ to be regarded as significant (see Breger
et al.\ 1999 for a more in-depth discussion of this criterion). The noise
level was calculated as the average amplitude in a 5 \cd interval centred
on the frequency of interest.

We repeated the prewhitening procedure with the $u$ and $v$ data
independently and obtained the same frequencies within the observational 
errors. We then determined final values for the detected frequencies by 
averaging the values from the individual filters, weighted by their $S/N$. 
The pulsational amplitudes were then recomputed with those frequencies; 
the result is listed in Table 2.

\begin{table}
\caption[]{Multifrequency solution for our time-resolved photometry of
$\nu$~Eri. Formal error estimates (following Montgomery \& O'Donoghue
1999) for the independent frequencies range from $\pm$ 0.00001 \cd\,for
$f_1$ to $\pm$ 0.00035 \cd\,for $f_8$. Formal errors on the amplitudes are
$\pm$ 0.2 mmag in $u$ and $\pm$ 0.1 mmag in $v$ and $y$. The S/N ratio 
quoted is for the $y$ filter data.}
\begin{center}
\scriptsize
\begin{tabular}{llcccc}
\hline
ID & Freq. & $u$ Ampl. & $v$ Ampl. & $y$ Ampl. & $S/N$ \\
 & (\cd) & (mmag) & (mmag) & (mmag) & \\
\hline
$f_1$ & 5.76327 & 73.5 & 41.0 & 36.9 & 137.0 \\
$f_3$ & 5.62006 & 34.6 & 23.9 & 22.7 & 83.6\\
$f_4$ & 5.63716 & 32.2 & 22.4 & 21.0 & 77.7\\
$f_2$ & 5.65393 & 37.9 & 26.4 & 25.1 & 92.6\\
$f_6$ & 6.24408 & 3.9 & 2.5 & 2.6 & 9.8 \\
$f_7$ & 6.26205 & 2.9 & 1.9 & 1.8 & 6.8\\
$f_8$ & 7.19994 & 1.3 & 0.9 & 1.1 & 4.3\\
$f_5$ & 7.89780 & 4.3 & 3.1 & 3.0 & 11.7\\
$f_A$ & 0.43218 & 5.5 & 3.2 & 3.2 & 7.1\\
$f_2+f_3$ & 11.27399 & 2.8 & 1.7 & 1.4 & 6.4 \\
$f_1+f_3$ & 11.38333 & 11.1 & 7.9 & 7.5 & 34.9\\
$f_1+f_4$ & 11.40043 & 10.9 & 7.7 & 7.1 & 33.2\\
$f_1+f_2$ & 11.41720 & 12.6 & 9.0 & 8.4 & 39.3\\
2$f_1$ & 11.52654 & 4.5 & 3.1 & 2.9 & 13.8\\
$f_1+f_5$ & 13.66107 & 1.6 & 1.2 & 1.1 & 6.4\\
$f_1+f_3+f_4$ & 17.02049 & 1.0 & 0.7 & 0.7 & 4.7 \\
$f_1+f_2+f3$ & 17.03726 & 4.4 & 3.1 & 2.6 & 16.9\\
$f_1+f_2+f_4$ & 17.05435 & 1.0 & 0.8 & 0.9 & 6.1\\
$f_1+2f_2$ & 17.07113 & 0.8 & 0.7 & 0.5 & 3.5\\
2$f_1+f_3$ & 17.14660 & 1.8 & 1.4 & 1.2 & 8.0\\
2$f_1+f_4$ & 17.16370 & 1.7 & 1.2 & 1.0 & 6.6\\
2$f_1+f_2$ & 17.18047 & 1.9 & 1.4 & 1.3 & 8.2\\
2$f_1+f_2+f_3$ & 22.80053 & 1.5 & 1.0 & 0.9 & 6.9\\
\hline
\end{tabular}
\normalsize
\end{center}
\end{table}

The residuals from this solution were searched for additional candidate
signals that may be intrinsic. We have first investigated the residuals in
the individual filters, then analysed the averaged residuals in the three
filters (whereby the $u$ data were divided by 1.5 to scale them to
amplitudes and rms scatter similar to that in the other two filters), and
finally applied statistical weights according to the recommendation by
Handler (2003). Some interesting features were found and are listed in
Table 3.

\begin{table}
\caption[]{Possible further signals. The data set in which they attained 
highest $S/N$ are indicated.}
\begin{flushleft}
%\scriptsize
\begin{tabular}{lll}
\hline
ID & Frequency & $S/N$ \\
 & (\cd) & \\
\hline
 & 0.2543 & 4.0 ($v$) \\
 & 6.221 & 3.6 ($uvy$, weighted) \\
 & 7.252 & 3.4 ($uvy$, weighted) \\
 & 7.914 & 3.1 ($u$) \\
$f_1+2f_3$ & 17.0034 & 3.2 ($u$) \\
$f_1+2f_2+f_3$ & 22.6912 & 3.5 ($u$) \\
$3f_1+f_2+f_3$ & 28.5638 & 3.1 ($u$) \\
\hline
\end{tabular}
%\normalsize
\end{flushleft}
\end{table}

In this table, the signal at 0.254 \cd is formally significant in the $v$
filter data, and noticeable peaks are present at the same frequency in
both the $u$ and $y$ filter data. However, we find the evidence from all
the data sets taken together not sufficiently convincing to claim reality
for this peak. Similar comments apply to the other signals listed in Table
3. We would however like to point out that the variations near 6--8 \cd
may all be components of multiplets that include detected modes. The
signals at frequencies higher than 17 \cd all coincide with combinations
of detected modes.

\subsection{The comparison stars}

We still have to analyse the light curves of $\mu$~Eri$- \xi$~Eri. To
this end, we computed the amplitude spectrum of these data and show them
in the upper panel of Fig.\ 3. One peak stands out; prewhitening it leaves
strong evidence for further variability of this star (Fig.\ 3, second
panel), but no more periodicities can be detected.

\begin{figure}
\includegraphics[width=84mm,viewport=00 05 252 301]{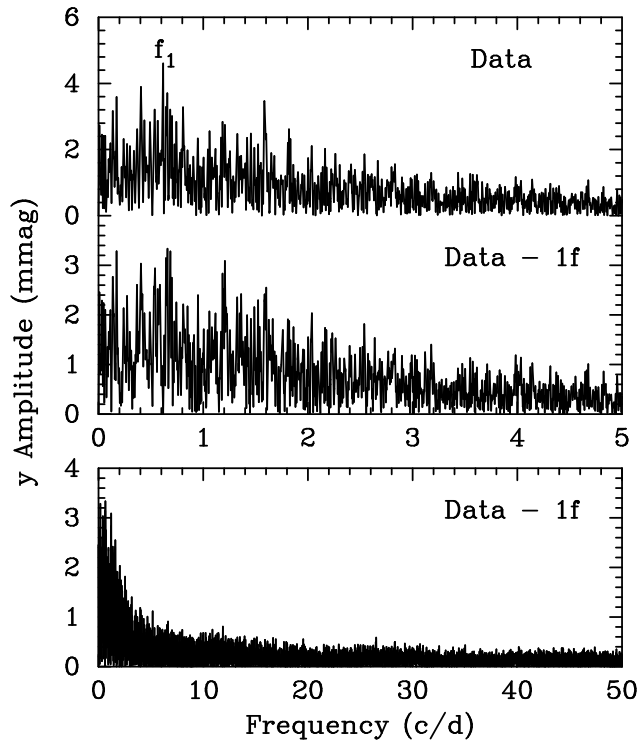}
\caption[]{Upper panel: amplitude spectrum of ($\mu$~Eri-$\xi$~Eri) in the
$y$ filter. Second panel: residual amplitude spectrum after prewhitening
$f_1$. Lower panel: residual amplitude spectrum of ($\mu$~Eri -$\xi$~Eri)
out to the Nyquist frequency; no further variations are detected.}
\end{figure}

The single periodicity that may be present in the light curves of
$\mu$~Eri has a frequency of 0.61638 $\pm$ 0.0005 \cd and $uvy$ amplitudes
of 10.2, 6.4 and 5.0 mmag, respectively. It is statistically significant
with S/N ratios between 6.3 and 4.4 in the different filters, but it is
not clear if its frequency and amplitude were constant throughout the
observing window. The residual amplitude spectrum after prewhitening this
signal still shows a very strong $1/f$ component and indicates that the
variability of $\mu$~Eri is complicated. However, further analyses of the
differential $\mu$~Eri$- \xi$~Eri light curves, such as searching for
variations with nonsinusoidal pulse shapes using the residualgram method
(Martinez \& Koen 1994) or by folding the data with the orbital period,
all remained inconclusive.

On the other hand, we noticed that the amplitude spectrum of the
de-trended comparison star magnitude differences (with a low-order
polynomial fitted to each night of data to remove the slow variability of
$\mu$~Eri) has the highest peak at 10.873 \cd in all the filters. It is
even significant with a $S/N=5.2$ in the $v$ filter data, where it reaches
an amplitude of 0.6 mmag.

To examine whether this peak is real and if so, from what star it
originates, we computed the differential ($\nu$~Eri $-\mu$~Eri) and ($\nu$
Eri$-\xi$~Eri) light curves. We then prewhitened the frequency solution
from Table 2 from these data and computed the residual amplitude spectra.  
We also examined the ($\xi$~Eri -- HD 29227) data from OSN for this
purpose. Unfortunately, these tests were not fully conclusive because the
noise level in these amplitude spectra is higher than in the ones of the
de-trended differential comparison star magnitudes only. All we can say is
that if real, the 10.873 \cd variation is more likely to originate from
$\xi$~Eri. In any case, our assumption that $\xi$~Eri is photometrically
constant does not affect our analysis of $\nu$~Eri.

\subsection{Re-analysis of literature data}

We have reanalysed the photometric measurements by van Hoof (1961) and
Kubiak \& Seggewiss (1991). The first data set was retrieved from the IAU
archives (as deposited by Cuypers \& Goossens 1981) and consists of two
seasons of ultraviolet $U$ and one season of yellow $Y$ measurements (the
$Y$ bandpass is identical to Johnson $V$, see Lyng\aa ~1959). The
frequency analysis was performed on the $U$ data as they are considerably
more numerous, whereas we determined only the amplitudes in the $Y$ data
with frequencies fixed to the values derived from the $U$ measurements.

We homogenised these data by averaging them into 7-minute bins and by
removing poor data, sometimes whole nights. The frequency analysis of the
resulting $U$ data set revealed the presence of frequencies $f_1$ to $f_4$
as well as the sum frequencies of $f_1$ with the $f_2$, $f_3$, $f_4$
triplet. The amplitude spectrum after prewhitening the corresponding
multifrequency fit shows a strong increase of noise that precludes the
detection of further signals. We note that neither the low 0.432 \cd
variation of $\nu$~Eri nor the suspected 0.616 \cd periodicity of
$\mu$~Eri could be detected.

To enable a search for further signals known from our analysis in those
data, we determined the zeropoints of the residual light curves of the
individual nights and subtracted them from the data. Fourier analysis of
this modified data set allowed the detection of four more signals
established in our measurements. Their frequencies and amplitudes
recovered in van Hoof's $U$ data are listed in Table 4 together with the
corresponding $Y$ amplitudes.

\begin{table}
\caption[]{Multifrequency solution for the $U$ and $Y$ data of van Hoof
(1961). The identifications of the signals are the same as in Table 2.
Formal error estimates (Montgomery \& O'Donoghue 1999) for the independent
frequencies range from $\pm$ 0.00001 \cd\,for $f_1$ to $\pm$ 0.00017
\cd\,for $f_6$. Formal errors on the $U$ amplitudes are $\pm$ 0.3 mmag.  
However, the real errors are believed to be higher because of the 
zeropoint adjustments we made. The formal uncertainty on the $Y$
amplitudes is $\pm$ 0.8 mmag.}
\begin{center}
\scriptsize
\begin{tabular}{llccc}
\hline
ID & Freq. & $U$ Ampl. & $Y$ Ampl. & $S/N$ \\
 & (\cd) & (mmag) & (mmag) & \\
\hline
$f_1$ & 5.76345 & 52.8 & 28.9 & 78.2 \\
$f_3$ & 5.62018 & 25.5 & 20.1 & 37.7\\
$f_4$ & 5.63738 & 26.5 & 19.5 & 39.2\\
$f_2$ & 5.65385 & 27.9 & 20.8 & 41.3\\
$f_6$ & 6.24417 & 2.7 & 2.1 & 4.0 \\
$f_7$ & 6.26227 & 3.7 & 4.0 & 5.6\\
$f_5$ & 7.89830 & 2.1 & 2.0 & 3.1\\
$f_1+f_3$ & 11.38363 & 8.1 & 6.2 & 13.0\\
$f_1+f_4$ & 11.40083 & 8.1 & 5.5 & 13.0\\
$f_1+f_2$ & 11.41730 & 9.2 & 7.0 & 14.7\\
2$f_1$ & 11.52690 & 2.8 & 3.8 & 4.4\\
\hline
\end{tabular}
\normalsize
\end{center}
\end{table}

The photometric measurements by Kubiak \& Seggewiss (1991) consist of two
runs of 17.1 and 5.1 days time base, respectively, separated by 4 years.
The $f_2$, $f_3$, $f_4$ triplet can therefore not be resolved in these
data, but $f_1$ can be separated from it and its $uvby$ amplitudes can be
estimated. We obtain amplitudes of $55\pm3$ mmag in $u$, $30\pm2$ mmag in
$v$, $28\pm2$ mmag in $b$, and $27\pm2$ mmag in $y$ for signal $f_1$. The
0.616 \cd variation of $\mu$ Eri may be present in these data.

\section{Discussion}

\subsection{The $\beta$ Cephei-type pulsation frequencies}

We have detected eight independent signals in the light curves of
$\nu$~Eri that are in the typical frequency domain for $\beta$ Cephei star
pulsation, i.e.\ they are pressure (p) and gravity (g) modes of low radial
order. We show the schematic amplitude spectrum composed of these modes
and of further suspected signals in Fig.\ 4.

\begin{figure}
\includegraphics[width=84mm,viewport=00 05 252 173]{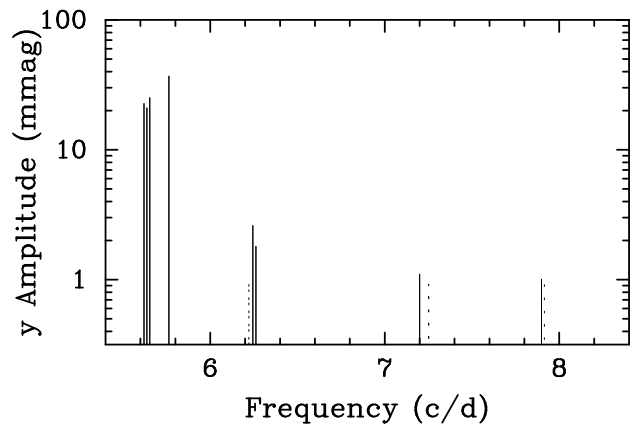}
\caption[]{Schematic amplitude spectrum of $\nu$~Eri. Solid lines 
represent detected modes, whereas the dashed lines indicate the positions 
of possible further signals.}
\end{figure}

This figure shows intriguing structures. Besides the known triplet of
frequencies near 5.64 \cd, there is another doublet near 6.24 \cd with a
suspected further component, and the two other signals near 7.2 and 7.9
\cd may also by parts of mode multiplets. We believe that these are signs
of rotational splitting of nonradial pulsation modes. If so, the rotation
period of $\nu$~Eri must be between 30 and 60 d, depending on the types of
mode we see (p and/or g modes).

The low-frequency triplet is asymmetric. Dziembowski \& Jerzykiewicz
(2003) calculated the asymmetry as $A_{\rm obs}=f_2+f_3-2f_4=-7.1\pm0.3
\times 10^{-4}$\cd (using the naming convention from Table 2) from
archival data, whereas our measurements indicate $A_{\rm obs}=-3.3\pm0.4
\times 10^{-4}$\cd. Comparing Tables 2 and 4, it appears that our value
for the triplet centroid frequency $f_4$ is less accurate than the
formal errors would suggest - which is plausible given that our
observational time base is only about 2.6 times the inverse triplet
splitting.

A comparison of Tables 2 and 4 allows another interesting conclusion: the
pulsational amplitudes of all modes of $\nu$~Eri seem to have increased
between van Hoof's and our measurements. Allowing for the different
wavelength passband of the archival $U$ measurements that also included a
silvered mirror (Lyng\aa ~1959) and our Str\"omgren $u$ data, an increase
in the pulsational amplitudes of about 20 per cent can be estimated. It is
possible that most of this increase has occurred in the last 15 -- 20
years as the $u$ amplitude of $f_1$ in the data by Kubiak \& Seggewiss
(1991) is also considerably smaller than in our data.

%An examination of the temporal stability of the pulsational frequencies 
%is difficult because of the large gaps between the data sets at our 
%disposal. However, the accuracy to which we could determine $f_1$ in van 
%Hoof's and our measurements allows us to determine a mean frequency 
%(5.7633036 \cd) over the total 45-year baseline. Evidence for a negative 
%period change is present; no similar analysis can be made for any other 
%pulsation mode.

\subsection{The combination signals}

The light curves of $\nu$~Eri are not perfectly sinusoidal (cf.\ Fig.\,1);
therefore combination signals result from our method of frequency
determination. It is not well known what the physical cause of combination 
frequencies is. Some of the most prominent hypotheses include simple 
light-curve distortions due to the pulsations propagating in a nonlinearly 
responding medium or independent pulsation modes excited by resonances.

In case of the light-curve distortion hypothesis, the amplitudes and
phases of the combination frequencies can be predicted. They also contain
some information about the medium that distorts the light curves (see Wu
2001 and references therein). In the simplest case, the amplitudes of the
combination signals scale directly with the product of the amplitudes of
the parent modes (e.g.\ see Garrido \& Rodr\I guez 1996).

As we do not have a pulsational mode identification available at this 
point, we cannot examine the two abovementioned hypotheses quantitatively, 
but some statements can still be made. For instance, simple light curve 
distortion should produce combination sum and difference frequencies of 
about the same amplitude. However, difference frequencies are completely 
absent in our multifrequency solution (Table 2) although the strongest of 
these signals should be easily detectable in our data.

The amplitude variations we reported before can also be examined. We note
that the amplitude of the sum frequencies of $f_1$ with the $f_2$, $f_3$,
$f_4$ triplet increased by about 20 per cent, which is the same as the
increase of the individual amplitudes of the parent modes, but less than
expected for the simple light curve distortion hypothesis.

The absence of the difference frequencies also poses a problem for the
resonant mode coupling hypothesis. As the stellar eigenmode spectrum is
much denser at low frequencies than at high frequencies due to the
presence of many gravity modes, one would expect to see many more
difference frequencies in case of mode coupling, which is not the case.

\subsection{The low-frequency variation}

We found a signal of 0.43218 \cd in the light curves of $\nu$~Eri and
evidence for other periodic low-frequency signals. Such variability is
an order of magnitude slower than $\beta$ Cephei-type pulsation. These
variations are not due to the slowly variable comparison star $\mu$~Eri:
they occur at frequencies different from the dominant variation of the
latter star and we have taken care that its variability does not affect
our analysis of $\nu$~Eri in our reduction procedures. They are therefore
present in our light curves of $\nu$~Eri.

Although this slow variability was detected in the measurements in all
three Str\"omgren filters used, we took special care in determining
whether these variations are intrinsic to the star. Consequently we 
computed amplitude spectra of the four largest homogeneous subsets of data 
(i.e. those that used the same filters and detectors throughout the whole 
campaign and that spanned a time base longer than 70 days) in all three 
filters. We found a peak at 0.432 \cd present in all these amplitude 
spectra, and we are therefore sure it is not an artifact of the observing 
or reduction procedures; it is due to intrinsic variations of $\nu$~Eri.

Unfortunately, there is some doubt as to whether this is an independent
frequency or not. A possible high-order combination frequency
($3f_1-3f_3$) would be located only 0.0026 \cd away from $f_A$, which is
much larger than the formal error estimate of $\pm 0.0001$ \cd for the
frequency uncertainty of $f_A$, but smaller than the 0.0063 \cd frequency
resolution of our data. It would be surprising if such a high-order
difference frequency were present in our data when there is no evidence
for lower-order ones. We thus suspect that $f_A$ is an independent
frequency, but only new measurements increasing the time base of our data
set will allow us to answer this question unambiguously.

If $f_A$ were an independent periodicity, what would be the physical
reason for such a variability? As argued above, the rotation period of
$\nu$~Eri must be between 30 and 60 d; the observed 2.3-d period can
therefore not be connected to rotation. $\nu$~Eri is also not known to be
a binary; extensive radial velocity studies are available and no
variability except that due to the short-period pulsations has ever been
reported. In addition, the wavelength dependence of the amplitude of the
0.432 \cd signal excludes a pure geometric origin of this variability.

Hence, the slow variations are probably due to high-order g-mode
pulsations of the star, which is also consistent with the colour
amplitudes. We note that Jerzykiewicz (1993) also detected a low-frequency
variation in his light curves of the $\beta$~Cephei star 16 (EN) Lac,
which he interpreted to be either due to a pair of spots - or due to
g-mode pulsation.

\subsection{The variability of $\mu$ Eri}

In Sect.\ 3.1 we reported that one of our comparison stars, $\mu$ Eri,
shows rather complex light variations with a time scale of about 1.6 days.  
To examine their origin, we first determined its position in the HR
diagram. As a start, we retrieved the standard Str\"omgren and Geneva
colours of $\mu$ Eri from the Lausanne-Geneva data base ({\tt
http://obswww.unige.ch/gcpd/gcpd.html}).

The star's Geneva colours then imply $T_{\rm eff}=15670\pm100$ K
according to the calibrations by K\"unzli et al. (1997). The star's
HIPPARCOS parallax (ESA 1997), combined with a reddening correction of
$A_{\rm V}= 0.06$ determined from its Str\"omgren colours and Crawford's 
(1978) calibration, bolometric corrections by Flower (1996) and Drilling 
\& Landolt (2000) result in $M_{\rm bol}=-3.6\pm 0.4$. We can thus place 
$\mu$ Eri in the theoretical HR diagram (Fig.\ 5).

\begin{figure}
\includegraphics[width=84mm,viewport=00 05 268 202]{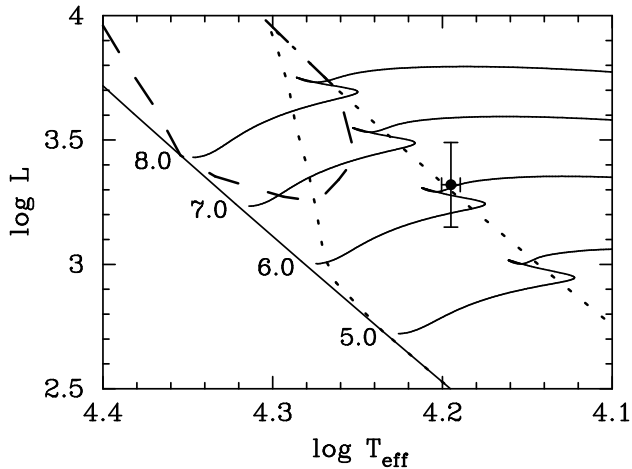}
\caption[]{The position of $\mu$ Eri in the theoretical HR diagram. Some
stellar model evolutionary tracks labelled with their masses (full lines)  
for $v \sin i_{\rm ZAMS}=200$\,km/s are included. The theoretical borders
of the $\beta$ Cephei and SPB star instability strips (Pamyatnykh 1999,
thick dashed and dotted lines) are included for comparison. The
evolutionary tracks are somewhat shifted to the ZAMS (full diagonal line)  
and to the instability strip boundaries because the latter do not include
rotational effects.}
\end{figure}

Interestingly, $\mu$ Eri seems to be an object just at or shortly after
the end of its main sequence life. It is located within the instability
strip of the slowly pulsating B (SPB) stars, and the time scale (cf.\ De
Cat \& Aerts 2002, Pamyatnykh 2002) and complexity of its variability as
well as the colour dependence of its amplitudes on wavelength are
consistent with it being an SPB star. $\mu$ Eri may have many pulsation
modes excited which are so closely spaced in frequency that we cannot
resolve them with the time base of our measurements.

However, pulsation is not the only possible physical cause of the light
variations of $\mu$ Eri. For the 1.622 d time scale of the variability of
the star to be due to rotational effects, an equatorial rotational
velocity of 193 km/s is required. The published estimates of the projected
rotational velocity $v \sin i$ of $\mu$ Eri range from 150 km/s (Abt et
al.\ 2000) to 190 km/s (Bernacca \& Perinotto 1970), which is consistent
with that constraint. A double-wave light variation with twice that period
would however be in conflict with the measured $v \sin i$. The complexity
of the light variations of $\mu$ Eri argues against a rotational origin.

The same argument can be used against an interpretation of $\mu$ Eri's
variability in terms of binarity. In addition (as argued above for the
long period detected for $\nu$ Eri), the wavelength dependence of the
colour amplitudes in our measurements suggests that a pure geometric
origin of this variability is unlikely. Most importantly, however, the
7.3-d spectroscopic binary period of $\mu$ Eri is quite different from the
observed time scale of its light variability or from an integral multiple
of it.

Hence, we cannot unambiguously determine the cause of the variability of 
$\mu$ Eri. A high-resolution, high signal-to-noise spectroscopic study 
will probably allow to distinguish between the two viable hypotheses, 
pulsation or rotational modulation.

\section{Conclusions}

Our photometric multisite campaign of the $\beta$ Cephei star $\nu$~Eri
resulted in the largest single data set ever obtained for such a pulsator.  
The frequency analysis of these measurements revealed the presence of
eight independent pulsation modes, which is the most ever detected for
such a star. These are normal $\beta$ Cephei-type variations (p and g
modes of low radial order), but one additional signal may be a high-order
gravity-mode pulsation. $\nu$~Eri could therefore be not only a $\beta$
Cephei star, but also an SPB star. It would thus the second example of a
star exhibiting two different types of pulsation. The high-order g-modes
of the first such star, HD 209295, are however believed to be triggered by
tidal effects (Handler et al.\ 2002). As we have no evidence for binarity
of $\nu$~Eri, it may therefore be the first star in which two types of
pulsation with time scales more than an order of magnitude different are
intrinsically excited.

The $\beta$ Cephei-type pulsation frequencies show some regular
structures, i.e.\ some are contained in multiplets that may be due to
nonradial m-mode splitting. If so, the rotation period of $\nu$~Eri is
between 30 -- 60 d, depending on the type of modes in these multiplets.  
These multiplets are a strong constraint for pulsational mode
identification, which will be the subject of a future paper. In any case,
the low-order p and g mode spectrum of $\nu$~Eri as reported here is
suitable for detailed seismic modelling.

\section*{ACKNOWLEDGEMENTS}

This work has been supported by the Austrian Fonds zur F\"orderung der
wissenschaftlichen Forschung under grant R12-N02. MJ's participation in
the campaign was partly supported by KBN grant 5P03D01420. MJ would also
like to acknowledge a generous allotment of telescope time and the
hospitality of Lowell Observatory.

We thank the following students of the University of Hawaii for their
assistance during some of the observations: Dan Bolton, Alexandre Bouquin,
Josh Bryant, Thelma Burgos, Thomas Chun, Alexis Giannoulis, Marcus
Lambert, Amanda Leonard, Alex MacIver, Danielle Palmese, Jennale Peacock,
David Plant, Ben Pollard, Sunny Stewart, Dylan Terry and Brian Thomas.

GH wishes to express his thanks to Lou Boyd and Peter Reegen for their
efforts in maintaining and controlling the Fairborn APTs, to Marcin Kubiak
for supplying his archival measurements of $\nu$~Eri, to Jan Cuypers for
helpful information on the passbands of the archival data, and to Conny
Aerts and Anamarija Stankov for comments on a draft version of this paper.

\bsp


\begin{thebibliography}{99}

\bibitem[]{}Abt, H. A., Levato H., Grosso M., 2002, ApJ 573, 359

\bibitem[]{}Aerts C., Waelkens C., de Pauw M., 1994, A\&A 286, 136

\bibitem[]{}Aerts C., Thoul A., Daszynska J., Scuflaire R., Waelkens C.,
Dupret M. A., Niemczura E., Noels A., 2003, Science 300, 1926

\bibitem[]{}Balona L. A., 2000, in {\it Delta Scuti and Related Stars},
ed.\ M. Breger \& M. H. Montgomery, ASP Conf. Ser. Vol. 210, p. 170

\bibitem[]{}Bernacca P. L., Perinotto M., 1970, Contr. Oss. Astrof. Padova
in Asiago, 239, 1B

\bibitem[]{}Breger, M., et al., 1999, A\&A 349, 225

\bibitem[]{}Crawford D. L., 1978, AJ 83, 48

\bibitem[]{}Cuypers J., Goossens M., 1981, A\&AS 45, 487

\bibitem[]{}Cuypers J., Aerts C., Buzasi D., Catanzarite J., Conrow T.,
Laher R., 2002, A\&A 392, 599

\bibitem[]{}Daszy{\'n}ska-Daszkiewicz J., Dziembowski W. A.,
Pamyatnykh A. A., Goupil M.-J., 2002, A\&A 392, 151

\bibitem[]{}De Cat P., Aerts C., 2002, A\&A 393, 365

\bibitem[]{}Drilling J. S., Landolt A. U., 2000, in {\it Allen's
Astrophysical Quantities}, 4$^{\rm th}$ edition, ed. A. N. Cox, Springer
Verlag, p.~392

\bibitem[]{}Dziembowski W. A., Pamyatnykh A. A., 1991, A\&A 248, L11

\bibitem[]{}Dziembowski W. A., Jerzykiewicz M., 2003, in {\it
International Conference on magnetic fields in O, B and A stars}, ed. L.
A. Balona, H. F. Henrichs \& R. Medupe, ASP Conf. Ser. in press

\bibitem[]{}ESA, 1997, The {\it Hipparcos} and Tycho catalogues, ESA 
SP-1200

\bibitem[]{}Flower P. J., 1996, ApJ 469, 355

\bibitem[]{}Garrido R., Rodr\I guez E., 1996, MNRAS 281, 696

\bibitem[]{}Handler G., 2003, Baltic Astronomy 12, 253

\bibitem[]{}Handler G., Aerts C., 2002, Comm. Asteroseismology 142, 20

\bibitem[]{}Handler G., et al., 1996, A\&A 307, 529

\bibitem[]{}Handler G., et al., 1997, MNRAS 286, 303

\bibitem[]{}Handler G., et al., 2000, MNRAS 318, 511

\bibitem[]{}Handler G., et al., 2002, MNRAS 333, 262

\bibitem[]{}Handler G., Shobbrook R. R., Vuthela F. F., Balona L. A.,
Rodler F., Tshenye T., 2003, MNRAS 341, 1005

\bibitem[]{}Heynderickx D., Waelkens C., Smeyers P., 1994, A\&AS 105, 447

\bibitem[]{}Hill G., 1969, Publ.\ DAO Victoria, 13, 323

\bibitem[]{}van Hoof A., 1961, Z. Astrophys., 53, 106

\bibitem[]{}de Jager C., 1963, Bull. Astr. Inst. Netherlands 17, 1

\bibitem[]{}Jerzykiewicz M., 1978, Acta Astr. 28, 465

\bibitem[]{}Jerzykiewicz M., 1993, Acta Astr. 43, 13

\bibitem[]{}Koen C., Eyer L., 2002, MNRAS 331, 45

\bibitem[]{}Kubiak M., 1980, Acta Astr.\ 30, 41

\bibitem[]{}Kubiak M., Seggewiss W., 1991, Acta Astr.\ 41, 127

\bibitem[]{}K\"unzli M., North P., Kurucz R. L., Nicolet B., 1997, A\&AS
122, 51

\bibitem[]{}Lyng\aa ~G., 1959, Arkiv for Astronomi 2, 379

\bibitem[]{}Martinez P., Koen C., 1994, MNRAS 267, 1039

\bibitem[]{}Montgomery M. H., O'Donoghue D., 1999, Delta Scuti Star
Newsletter 13, 28 (University of Vienna)

\bibitem[]{}Moskalik P., Dziembowski W. A., 1992, A\&A 256, L5

\bibitem[]{}Nather R. E., Winget D. E., Clemens J. C., Hansen C. J.,
Hine B. P., 1990, ApJ 361, 309

\bibitem[]{}Pamyatnykh A. A., 1999, Acta Astr. 49, 119

\bibitem[]{}Pamyatnykh A. A., 2002, Comm. Asteroseismology 142, 10

\bibitem[]{}Pamyatnykh A. A., 2003, Ap\&SS 284, 97

\bibitem[]{}Sperl M., 1998, Master's Thesis, University of Vienna

\bibitem[]{}Stankov A., Handler G., Hempel M., Mittermayer P., 2002, MNRAS
336, 189

\bibitem[]{}Sterken C., et al., 1986, A\&AS 66, 11

\bibitem[]{}Wu Y., 2001, MNRAS 323, 248

\bibitem[]{}Zima W., et al., 2002, in {\it Radial and Nonradial Pulsations
as Probes of Stellar Physics}, ed. C. Aerts, T. R. Bedding \& J.
Christensen-Dalsgaard ASP Conf. Proc., Vol.\ 259, p.\ 598

\end{thebibliography}
\end{document}